\documentclass{aa}
\usepackage{graphicx,natbib}
\usepackage[varg]{txfonts}
\usepackage{color}

\newcommand{\Hm}{\langle B\rangle}
\newcommand{\Hz}{\langle B_z\rangle}

\newcommand{\Hav}{\Hm_{\rm av}}

\newcommand{\Hzrms}{\Hz_{\rm rms}}

\newcommand{\vsi}{v\,\sin i}
\newcommand{\Prot}{P_{\rm rot}}

\newcommand{\Zeeman}{\Delta\lambda_{\rm Z}}

\newcommand{\Feline}{Fe~{\sc ii}~$\lambda\,6149.2$}

\bibpunct[ ]{(}{)}{;}{a}{}{,}

\begin{document}

\title{The variation of the magnetic field of the Ap star HD~50169\\
  over its 29 year rotation period\thanks{Based on observations
    collected at the European 
  Southern Observatory under ESO Programmes 62.L-0867, 64.L-0443, and
  2100.D-5013; on observations collected  
at the 6-m telescope BTA of the Special Astrophysical Observatory of
the Russian Academy of Sciences; and on data products from
observations made with ESO Telescopes at the La Silla Paranal
Observatory under programmes
68.D-0254, 74.C-0102, 75.C-0234, 76.C-0073, 79.C-0170, 80.C-0032,
81.C-0034, and 82.C-0308.}}

\author{G.~Mathys\inst{1} 
  \and I.~I.~Romanyuk\inst{2,3}
  \and S. Hubrig\inst{4}
\and D.~O.~Kudryavtsev\inst{2}
\and \\J.~D.~Landstreet\inst{5,6}
\and M.~Sch\"oller\inst{7}
\and E.~A.~Semenko\inst{2}
\and I.~A.~Yakunin\inst{2}}

\institute{European Southern Observatory,
  Alonso de Cordova 3107, Vitacura, Santiago, Chile\\\email{gmathys@eso.org}
\and
Special Astrophysical Observatory, Russian Academy of Sciences,
Nizhnii Arkhyz, 369167 Russia
\and
Institute of Astronomy of the Russian Academy of Sciences, 48 Pyatnitskaya St, 119017 Moscow, Russia
\and
Leibniz-Institut für Astrophysik Potsdam (AIP), An der Sternwarte 16, 14482 Potsdam, Germany
\and
Department of Physics \& Astronomy, University of Western Ontario,
1151 Richmond Street, 
London, Ontario N6A 3K7, Canada 
\and
Armagh Observatory, College Hill, Armagh, BT61 9DG, Northern Ireland,
UK 
\and
European Southern Observatory, Karl-Schwarzschild-Str. 2, 85748 Garching bei M\"unchen, Germany}

\date{Received $\ldots$ / Accepted $\ldots$}

\titlerunning{The 29 year rotation period of HD~50169}

\authorrunning{G. Mathys et al.}

\abstract
{The Ap stars that rotate extremely slowly, with periods of decades to
  centuries, represent one of the keys to the understanding of the
  processes leading to the differentiation of stellar rotation.} 
{We characterise the variations of the magnetic field of the Ap star
  HD~50169 and derive constraints about its structure.} 
{We combine published measurements of the mean longitudinal field
  $\Hz$ of HD~50169 with new determinations of this field moment from
  circular spectropolarimetry obtained at the 6-m telescope BTA of the
  Special Astrophysical  Observatory of the Russian Academy of
  Sciences. For the mean magnetic field modulus $\Hm$, literature data
  are complemented by the analysis of ESO spectra, both newly acquired
  and from the archive. Radial velocities are also obtained from
  these spectra.} 
{We present the first determination of the rotation period of
  HD~50169, $\Prot=29.04\pm0.82$\,y. HD~50169 is currently the
  longest-period Ap star for which magnetic field measurements have
  been obtained over more than a full cycle. The variation curves of
  both $\Hz$ and $\Hm$ have a significant degree of anharmonicity, and
  there is a definite phase shift between their respective extrema. We
  confirm that HD~50169 is a wide spectroscopic binary, refine its
  orbital elements, and suggest that the secondary is probably a dwarf
  star of spectral type M.} 
{The shapes and mutual phase shifts of the derived magnetic variation
  curves unquestionably indicate that the magnetic field of HD~50169
  is not 
  symmetric about an axis passing through its centre. Overall,
  HD~50169 appears similar to the bulk of the 
  long-period Ap stars.} 


\maketitle

\section{Introduction}
\label{sec:intro}
One of the most intriguing properties of the Ap stars is their slow
rotation, compared to the superficially normal main-sequence stars in
the same temperature range. In recent years, it has emerged that
several percent of the Ap stars have rotation periods exceeding one
month, and that the 
longest rotation periods reach several centuries
\citep{2017A&A...601A..14M}. Perhaps even more 
puzzling is the spread of the rotation periods of the Ap stars, as
some of them are as short as 0.5\,d. How period differentiation over
at least five orders of magnitude
is achieved in stars that are essentially at the same evolutionary
stage remains to be fully explained. \citet{2018arXiv181006106S}
recently argued that it is 
impossible for Ap stars to achieve sufficient braking once they are
formed, so that they must have acquired their slow rotation already
during the accretion phase, through a magnetic connection between the
accreting matter and the birth cloud. This is only the conclusion of the
latest attempt to build up a theory that accounts for the observed
distribution of the rotation periods of the Ap stars. But like all
previous similar attempts, the possibility of testing critically this
theory is strongly limited by the still very incomplete knowledge of
the long-period end of that distribution. 

In turn, this incompleteness arises in large part -- but definitely
not entirely -- from the finite time base over which relevant data
have been collected. The first determination of the rotation period of
an Ap star ($\alpha^2$~CVn = HD~112413, $\Prot=5.5$\,d) was achieved
just over one century ago \citep{1913AN....195..159B}, and hardly more
than 70 years have elapsed since the first detection of a
magnetic field in such a star \citep[78~Vir =
HD~118022,][]{1947ApJ...105..105B}. CS~Vir (= HD~125248) was the first
star in which the periodicity of the variations of the magnetic field
was established \citep{1951ApJ...114....1B}; the magnetic and spectral
variations were found to have the same period. By the end of the decade,
out of $\sim$70 Ap stars known to be magnetic,
\citet{1958ApJS....3..141B} identified only five as showing definitely
periodic magnetic variations. By that time, however,
\citet{1956PASP...68...92D} had convincingly shown that the Oblique
Rotator Model \citep{1950MNRAS.110..395S} could account for
the observed magnetic, spectroscopic and photometric variability of
the Ap stars. Eventually, 
\citet{1970stro.coll..254P} presented the definitive arguments
supporting this model. The only remaining issue was how the few Ap
stars with very long periods that were known at the time fit in the
context of this model. This triggered increased interest in these
stars. \citet{1970PASP...82..878P} himself compiled a list of stars
that might have long periods. He took advantage of the low $\vsi$ of
these stars to carry out a first systematic study of the mean magnetic
field modulus, discovering five new stars with resolved magnetically
split lines (in addition to the four that were previously known) and
estimating ``surface 
fields'' (or upper limits) for 21 more \citep{1971ApJ...164..309P}. 

In other words, the dedicated investigation of the longest variation
periods of Ap stars started less than 50 years ago. Highlights of the
first two decades of this effort include the works of
\cite{1975ApJ...202..127W}, \cite{1984A&A...132..291H}, and
\cite{1988A&A...199..299R}. However, 
the overall picture remained patchy. Photoelectric photometry, the
tool of choice for the determination of Ap star periods of up to a
couple of weeks, proved much less suited to the study of
slower variations, because of its limitations with respect to
long-term stability and reproducibility. The discovery of new Ap stars
with resolved magnetically split lines \citep{1990A&A...232..151M},
and the realisation that there were many more such stars to be found
\citep{1992A&A...256..169M}, triggered a systematic search, and 
extensive follow-up monitoring, to identify those stars and
characterise their magnetic fields
\citep{1997A&AS..123..353M,2017A&A...601A..14M}. As a result, many new
very slowly rotating stars were found.

The milestones identified above set upper limits to the time base over
which magnetic field measurements of the Ap stars were obtained:
$\sim$70\,y since the first magnetic stars were discovered, $\sim$50\,y
since the interest of the stars with the longest periods was
recognised, and $\sim$25\,y since stars with very long periods
started to be discovered in statistically significant numbers. The
majority of the longest period stars have only been identified more
recently and observed for less than 25 years, so that their variations
could not be fully characterised yet. But even for the long period star
that has been followed for the longest time, $\gamma$~Equ  (=
HD~201601), one of the first half dozen Ap stars in which a magnetic
field was detected, for which the first longitudinal field
measurements were obtained in 1946 \citep{1948ApJ...108..191B}, a full
rotation cycle has not been observed yet. The most recent estimate of the
lowest plausible value of the period for this star puts it at 97\,y
\citep{2016MNRAS.455.2567B}, and the possibility that it is even
longer cannot be ruled out. Existing $\Hz$ measurements still cover
at most only $\sim$70\% of the full rotation cycle of this
star. Incidentally, the first measurement of its mean magnetic field
modulus from the magnetic splitting of a spectral line
was obtained less than 30 years ago
\citep{1990A&A...232..151M}, so that the current coverage of the $\Hm$
variation curve is less than one third of the rotation period. In most
cases, the phase coverage achieved so far is even more incomplete. For
instance, \citet{2018MNRAS.477.3791H} recently proposed a tentative
value of 188\,y for the rotation period of HD~101065 (Przybylski's
star), by extrapolation of magnetic observations that would span less
than one fourth of a rotation cycle. 
 
These considerations illustrate the fundamental limitations to our ability
to characterise the most slowly rotating Ap stars and emphasise the
value of any new addition to the small group of very long period stars
whose magnetic variations have been observed over a full rotation
cycle. Prior to this study, the five Ap stars with the longest periods
for which complete magnetic variation curves had been obtained were,
in order of decreasing period length:
HD~9996 \citep[= HR~465, $\Prot=7937$\,d,][]{2014AstBu..69..315M},
HD~94660 \citep[= HR~4263, $\Prot=2800$\,d,][]{2017A&A...601A..14M},
HD~187474 \citep[= HR~7552, $\Prot=2345$\,d,][]{1991A&AS...89..121M},
HD~59435 \citep[= BD~$-$8~1937, $\Prot=1360$\,d,][]{1999A&A...347..164W},
and HD~18078 \citep[= BD~+55~726, $\Prot=1358$\,d,][]{2016A&A...586A..85M}.

HD~50169 (=\,BD~$-1$~1414) is an A3p SrCrEu star
\citep{2009A&A...498..961R} in which the presence of a strong magnetic
field was first reported by \citet{1958ApJS....3..141B} from
spectroscopic observations in circular
polarisation. \citet{1971ApJ...164..309P} noted its very low $\vsi$
($\null<10$\,km\,s$^{-1}$) and obtained a first estimate of its mean
magnetic field modulus (which he referred to as mean surface magnetic
field), 5.6\,kG, from consideration of the differential broadening of
spectral lines with Zeeman patterns of different widths. Following the
discovery that the \Feline\ line is resolved into its magnetically
split components in the spectrum of HD~50169
\citep{1992A&A...256..169M}, more magnetic measurements started to be
acquired, both of the mean longitudinal field $\Hz$
\citep{1997A&AS..124..475M,2014AstBu..69..427R,2017A&A...601A..14M}
and of the mean field modulus $\Hm$ \citep{1997A&AS..123..353M,
  2017A&A...601A..14M}. Combining all the available measurements,
\citet{2017A&A...601A..14M} concluded that the rotation period of the
star must be much longer than 7.5\,y. However, he also argued that
this period should likely be shorter than 40\,y.

Here we present additional determinations of $\Hz$ and $\Hm$, based on
both new dedicated observations and archive spectra, and we use
them to derive for the first time the value of the rotation period of
HD~50169. The observational data and their analysis are described in
Sect.~\ref{sec:obs}, and the determination of the stellar rotation
period is presented in Sect.~\ref{sec:per}. In
Sect.~\ref{sec:magfield}, we derive constraints on the geometrical
structure of the magnetic field, and in Sect.~\ref{sec:bin} we refine
the determination of the orbital elements of the HD~50169 binary
system. Finally, we draw conclusions and discuss future prospects
(Sect.~\ref{sec:concl}). 

\section{Observations and data analysis}
\label{sec:obs}
\subsection{Mean magnetic field modulus}
\label{sec:Hm}
The mean magnetic field modulus $\Hm$ is the average over the visible
stellar disk of the modulus of the field vector, weighted by the local
emergent line intensity. The following published measurements of this
field moment were used in this study:
\begin{itemize}
\item 13 measurements from \citet{1997A&AS..123..353M},
\item and 8 measurements from \citet{2017A&A...601A..14M}.
\end{itemize}
These measurements were obtained with four different instrumental
configurations. For the sake of simplicity, we use the same symbols as
\citet{2017A&A...601A..14M} to identify those configurations in
Fig.~\ref{fig:bmcurve}.

\begin{table}
\caption{Mean magnetic field modulus and radial velocity measurements.}
\label{tab:Hm}
\centering
\begin{tabular}{lrcl}
\hline\hline\\[-4pt]
  \multicolumn{1}{c}{JD}&\multicolumn{1}{c}{$\Hm$}&\multicolumn{1}{c}{HRV}&Reference\\
  &\multicolumn{1}{c}{(G)}&\multicolumn{1}{c}{(km\,s$^{-1}$)}&\\[4pt]
\hline\\[-4pt]
 2448323.540&4448&    &\protect{\citet{1997A&AS..123..353M}}\\          
 2448914.842&4599&11.7&\protect{\citet{1997A&AS..123..353M}}\\          
 2448933.724&4612&12.3&\protect{\citet{1997A&AS..123..353M}}\\
 2448963.658&4654&12.0&\protect{\citet{1997A&AS..123..353M}}\\
 2448991.590&4723&11.8&\protect{\citet{1997A&AS..123..353M}}\\
 2449018.559&4694&11.6&\protect{\citet{1997A&AS..123..353M}}\\
 2449079.506&4681&12.5&\protect{\citet{1997A&AS..123..353M}}\\
 2449398.556&4885&12.7&\protect{\citet{1997A&AS..123..353M}}\\
 2449419.556&4906&12.6&\protect{\citet{1997A&AS..123..353M}}\\
 2449436.521&4945&12.8&\protect{\citet{1997A&AS..123..353M}}\\
 2449709.638&5038&13.6&\protect{\citet{1997A&AS..123..353M}}\\
 2449752.556&5120&13.6&\protect{\citet{1997A&AS..123..353M}}\\
 2449800.527&4964&13.7&\protect{\citet{1997A&AS..123..353M}}\\
 2450057.065&5266&15.6&\protect{\citet{2017A&A...601A..14M}}\\
 2450084.633&5220&15.4&\protect{\citet{2017A&A...601A..14M}}\\
 2450110.622&5220&15.3&\protect{\citet{2017A&A...601A..14M}}\\
 2450132.541&5285&15.4&\protect{\citet{2017A&A...601A..14M}}\\
 2450149.522&5354&15.4&\protect{\citet{2017A&A...601A..14M}}\\
 2450427.744&5470&13.0&\protect{\citet{2017A&A...601A..14M}}\\
 2450817.678&5724&12.4&\protect{\citet{2017A&A...601A..14M}}\\
 2451084.833&5812&12.3&\protect{\citet{2017A&A...601A..14M}}\\
 2451292.525&5954&13.0&This paper (CES VLC)\\
 2451600.552&6000&14.0&This paper (CES VLC)\\
 2452242.666&6099&12.1&This paper (UVES)\\
 2453333.866&5836&13.7&This paper (HARPS)\\
 2453463.501&5784&14.6&This paper (HARPS)\\
 2453464.499&5749&14.7&This paper (HARPS)\\
 2453714.671&5636&14.8&This paper (HARPS)\\
 2453716.643&5604&14.8&This paper (HARPS)\\
 2454222.500&5352&12.0&This paper (HARPS)\\
 2454223.513&5343&11.8&This paper (HARPS)\\
 2454224.526&5333&12.1&This paper (HARPS)\\
 2454336.919&5240&11.8&This paper (HARPS)\\
 2454338.916&5343&12.0&This paper (HARPS)\\
 2454441.684&5230&12.0&This paper (HARPS)\\
 2454442.741&5203&12.0&This paper (HARPS)\\
 2454443.698&5246&12.0&This paper (HARPS)\\
 2454443.770&5241&12.1&This paper (HARPS)\\
 2454543.499&5157&12.1&This paper (HARPS)\\
 2454544.638&5118&12.0&This paper (HARPS)\\
 2454545.513&5145&12.1&This paper (HARPS)\\
 2454717.908&5067&12.6&This paper (HARPS)\\
 2454718.913&5049&12.4&This paper (HARPS)\\
 2454865.698&5016&12.7&This paper (HARPS)\\
 2454866.626&4980&12.8&This paper (HARPS)\\
 2454868.707&4989&12.9&This paper (HARPS)\\
 2454869.717&4960&12.9&This paper (HARPS)\\
 2458075.841&4266&12.0&This paper (UVES)\\
 2458179.557&4287&12.2&This paper (UVES)\\[4pt]
\hline
\end{tabular}
\end{table}

Here, we present new $\Hm$ data at additional epochs, from the
analysis of the following high-resolution spectra recorded in natural
light:
\begin{itemize}
  \item 2 spectra recorded with the Very Long Camera (VLC) of the ESO
    Coud\'e Echelle Spectrograph (CES) fed by the ESO 3.6-m telescope
    \citep{2005CESVLC}. This configuration is quite different from
    those used by \citet{1997A&AS..123..353M} and
    \citet{2017A&A...601A..14M}, but the reduction process was very
    similar to the one applied for other fibre-fed CES configurations
    in these previous studies \citep[see Section 3 of][for
    details]{1997A&AS..123..353M}. 
    \item 23 spectra recorded with the High Accuracy Radial velocity
      Planet Searcher (HARPS) fed by the ESO 3.6-m telescope,
      retrieved from the  ESO Archive. 
      \item 3 spectra recorded with the Ultraviolet and Visible
        Echelle Spectrograph (UVES) fed by Unit Telescope 2 (UT2) of
        the ESO Very 
        Large Telescope (VLT). One of these spectra was retrieved from the
        ESO Archive, the other two were purpose-made DDT observations
        intended to constrain the strength of the mean magnetic field
        modulus around the phase of its minimum.
      \end{itemize}
For the HARPS and UVES observations, we used science grade pipeline
processed data available from the ESO Archive. The only additional
processing that we carried out was a continuum normalisation of the
region ($\sim$100\,\AA\ wide) surrounding the \Feline\ diagnostic line. 

\begin{table}
\caption{Mean longitudinal magnetic field measurements.}
\label{tab:Hz}
\centering
\begin{tabular}{crrl}
\hline\hline\\[-4pt]
\multicolumn{1}{c}{JD}&\multicolumn{1}{c}{$\Hz$ (G)}&$\sigma_z$ (G)&Reference\\[4pt]
\hline\\[-4pt]
2434651.000 &   670&  64&\citet{1958ApJS....3..141B}\\
2435023.000 &   890&  49&\citet{1958ApJS....3..141B}\\
2435026.000 &   910&  55&\citet{1958ApJS....3..141B}\\
2435145.000 &  1040&  79&\citet{1958ApJS....3..141B}\\
2435560.000 &  1030&  73&\citet{1958ApJS....3..141B}\\
2435765.000 &  2120&  74&\citet{1958ApJS....3..141B}\\
2449026.637 &  1297&  74&\protect{\citet{1997A&AS..124..475M}}\\
2449830.504 &  1035&  70&\protect{\citet{2017A&A...601A..14M}}\\
2449974.857 &  1002&  85&\protect{\citet{2017A&A...601A..14M}}\\
2450039.684 &   763&  39&\protect{\citet{2017A&A...601A..14M}}\\
2450111.554 &   911&  64&\protect{\citet{2017A&A...601A..14M}}\\
2450183.543 &   891&  52&\protect{\citet{2017A&A...601A..14M}}\\
2450497.591 &   471&  75&\protect{\citet{2017A&A...601A..14M}}\\
2450784.745 &   188&  60&\protect{\citet{2017A&A...601A..14M}}\\
2450832.727 &   115&  94&\protect{\citet{2017A&A...601A..14M}}\\
2452625.477 & $-$1536&  83&This paper\\
2452689.320 & $-$1464&  76&This paper\\
2452690.288 & $-$1630&  78&This paper\\
2452918.564 & $-$1493& 106&This paper\\
2453273.567 & $-$1790&  80&This paper\\
2453365.521 & $-$1430&  60&This paper\\
2453666.525 & $-$1400&  50&This paper\\
2453867.496 & $-$1420&  60&This paper\\
2454016.545 & $-$1190&  50&This paper\\
2454162.346 & $-$1020&  50&\citet{2014AstBu..69..427R}\\
2455282.239 &   130&  50&This paper\\
2457382.452 &  1600&  50&This paper\\
2457414.380 &  1380&  15&This paper\\
2457740.400 &  1400&  27&This paper\\
2457764.452 &  1423&  20&This paper\\
2457829.291 &  1610&  23&This paper\\
2457830.339 &  1544&  25&This paper\\
2457860.205 &  1380&  30&This paper\\
2458008.437 &  1590&  20&This paper\\
2458009.438 &  1540&  20&This paper\\
2458068.566 &  1410&  20&This paper\\
2458117.306 &  1510&  20&This paper\\
2458125.406 &  1470&  20&This paper\\
2458151.427 &  1620&  20&This paper\\[4pt]
\hline
\end{tabular}
\end{table}

This line is resolved in its two magnetically split components
in all the spectra that we analysed. We measured the wavelength
separation of the components to determine the mean magnetic field
modulus $\Hm$ at the corresponding epochs, by application of the formula:
\begin{equation}
\lambda_{\rm r}-\lambda_{\rm b}=g\,\Zeeman\,\Hm\,.
\label{eq:Hm}
\end{equation}
In this equation, $\lambda_{\rm r}$ and $\lambda_{\rm b}$ are,
respectively, the wavelengths of the red and blue split line
components; $g$ is the Land\'e factor of the split level of the
transition ($g=2.70$; \citealt{1985aeli.book.....S}); 
$\Zeeman=k\,\lambda_0^2$, with
$k=4.67\,10^{-13}$\,\AA$^{-1}$\,G$^{-1}$; $\lambda_0=6149.258$\,\AA\
is the nominal wavelength of the considered transition.

There is no reason to expect the uncertainty of the new determinations
of $\Hm$ presented here to be significantly different from that of the
previous measurements of \citet{1997A&AS..123..353M} and
\citet{2017A&A...601A..14M}. We shall see in
Sect.~\ref{sec:per} that the scatter of the measurements about
the variation curve of the mean magnetic field modulus is fully
consistent with this adopted value of the uncertainty,
30\,G. Incidentally, the $\Hm$ measurements in HD~50169 are among the
most precise that can be obtained for any Ap star with resolved
magnetically split lines, thanks to the fact that the split components
of the \Feline\ line are among the sharpest, best resolved, and least
blended of all, as can be seen in Figures 2 to 4 of \citet{1997A&AS..123..353M}.

The 49 values of the mean magnetic field modulus obtained in that way are
presented in Table~\ref{tab:Hm}. For the convenience of the reader,
this table also includes the previously published measurements. The
columns give, in order, the Julian Date of the observation, the value
$\Hm$ of the mean magnetic field modulus, the heliocentric radial
velocity, and the source of the measurement. Note that the radial
velocities determined by \citet{1997A&AS..123..353M} were
published by \citet{2017A&A...601A..14M}. For the new
determinations from this paper, the instrument that was used is
specified; for published measurements, this information is available
in the cited reference. The date that is given is the Heliocentric
Julian Date (or Barycentric, for the HARPS observations) of mid-exposure,
except for the two measurements based on CES VLC spectra, for which
the headers were incomplete. For these two spectra, the listed
mid-exposure time is accurate to the minute; its uncertainty does not
have any significant impact on the present study given the much longer
timescales involved. The radial velocity could not be determined for
the first observation listed in this table, owing to the lack of
appropriate calibration. As mentioned elsewhere, the estimated
uncertainties of the field modulus and radial velocity values are,
respectively, 30\,G and 1\,km\,s$^{-1}$.

\subsection{Mean longitudinal magnetic field}
\label{sec:Hz}
The mean longitudinal magnetic field $\Hz$ is the average over the
visible hemisphere of the component of the magnetic vector along the
line of sight, weighted by the local emergent line intensity. The
following published measurements of this field moment were used in
this study: 
\begin{itemize}
\item 6 measurements from \citet{1958ApJS....3..141B},
\item 1 measurement from \citet{1997A&AS..124..475M},
\item 1 measurement from \citet{2014AstBu..69..427R},
\item and 8 measurements from \citet{2017A&A...601A..14M}. 
\end{itemize}

All these measurements were carried out through the analysis of medium
to high
spectral resolution observations of metal lines in circular
polarisation. Here they are complemented by additional $\Hz$
determinations obtained from observations of the same type, that is,
spectra of HD~50169 recorded at $R\simeq14\,500$ in both circular
polarisations with the Main Stellar Spectrograph of the 6-m telescope
BTA of the Special Astrophysical Observatory, on 23 nights spread from
December 2002 to February 2018. This is the same configuration as used
by \citet{2014AstBu..69..427R}. The instrumental configuration and the data
reduction procedure are as described in detail in this reference.

The mean longitudinal magnetic field was determined from the
wavelength shifts of a sample of spectral lines between 
the two circular polarisations in each of these spectra, by application of the formula:
\begin{equation}
\lambda_{\rm R}-\lambda_{\rm L}=2\,\bar g\,\Zeeman\,\Hz\,,
\label{eq:Hz}
\end{equation}
where $\lambda_{\rm R}$ (resp. $\lambda_{\rm L}$) is the wavelength of
the centre of gravity of the line in right (resp. left) circular
polarisation and $\bar g$ is the effective Land\'e factor of the
transition. $\Hz$ is determined through a 
least-squares fit of the measured values of $\lambda_{\rm
  R}-\lambda_{\rm L}$ by a function of the form given above. The standard error
$\sigma_z$ that is derived from that
least-squares analysis is used as an estimate of the uncertainty
affecting the obtained value of $\Hz$. 

The values of the mean longitudinal field obtained in that way are
presented in Table~\ref{tab:Hz}. For the convenience of the reader,
this table also includes the previously published measurements. The
columns give, in order, the Julian Date of the observation, the value
$\Hz$ of the mean longitudinal magnetic field and its uncertainty
$\sigma_z$, and the source of the measurement. The accuracy of the
date of observation is variable, ranging from the Heliocentric Julian
Date of mid-exposure \citep{1997A&AS..124..475M,2017A&A...601A..14M} 
to the mere date of observation, without any time specification
\citep{1958ApJS....3..141B}. The associated phase uncertainty is only
$10^{-5}$, which is completely negligible. In total, 39 $\Hz$
measurements are listed in Table~\ref{tab:Hz}. 

\subsection{Radial velocities}
\label{sec:RC}
Following \citet{2017A&A...601A..14M}, we computed the
unpolarised wavelength $\lambda_I$ of the Fe~{\sc ii}
$\lambda\,6149.2$ line by averaging the
wavelengths of its blue and red split components as measured for the
determination of the mean magnetic field modulus, as follows:
\begin{equation}
\lambda_I=(W_{\lambda,{\rm b}}\,\lambda_{\rm b}+W_{\lambda,{\rm
    r}}\,\lambda_{\rm r})/(W_{\lambda,{\rm b}}+W_{\lambda,{\rm r}})\,.
\label{eq:rv6149}
\end{equation}
Then, the wavelength difference $\lambda_I-\lambda_0$
was converted to a radial velocity value in the standard
manner.
The notations $W_{\lambda,{\rm b}}$ and $W_{\lambda,{\rm r}}$ refer
to the equivalent widths of the measured parts of each line
component. For more details, see \citet{2017A&A...601A..14M}.

We applied this approach to all the additional high-resolution spectra
recorded in natural light that we used to measure $\Hm$. The
arguments presented by \citet{2017A&A...601A..14M} in support of the
validity of this procedure are further strengthened by the
consideration that the CES~VLC, UVES and HARPS are instruments (or
instrumental configurations) that were designed to optimise radial velocity
determinations for exoplanet searches and studies.

The new radial velocity measurements of this study are listed in
Table~\ref{tab:Hm}. We adopt for them the same uncertainty as for the
similar measurements of \citet{2017A&A...601A..14M}. This is somewhat
arbitrary, but it seems justified since we did not take any
particular care to optimise the radial velocity determinations, and
the scatter of 
the new measurements about the revised orbital solution presented in
Sect.~\ref{sec:bin} is of the same order as the scatter of the data
from \citet{2017A&A...601A..14M}.

\section{Magnetic variability and rotation period}
\label{sec:per}
To determine the rotation  period of HD~50169, we fitted the
measurements of its mean longitudinal magnetic field by either a cosine
wave, or the superposition of a cosine wave and of its first
harmonic, progressively varying the period of these waves, in search
of the value of the period that minimises the reduced $\chi^2$ of the
fit. These fits are weighted by the inverse of the square of
  the uncertainties of the individual measurements. Through this
procedure, we concluded unambiguously that the 
rotation period of the star must be of the order of 10,800\,d. 

  The usage of the procedure described above is justified by the well
  documented observation that the variation curves of the magnetic
  field moments of Ap stars often closely resemble a cosine wave or
  the superposition of a cosine wave and of its first harmonic
  \citep[e.g.][and references therein]{2017A&A...601A..14M}. However,
  there is no physical reason why the variation curve of either $\Hz$
  or $\Hm$ should have {\em exactly\/} such a shape. The complexity of
  the geometrical structure of the Ap star magnetic fields \citep[a prominent
  example is 53~Cam = HD~65339, as shown by][]{2004A&A...414..613K}
  implies 
  that whenever sufficiently precise determinations of a field moment
  with good enough phase sampling are available \citep[neither of
  which is fulfilled by the $\Hm$ measurements of HD~65339
  of][]{2017A&A...601A..14M}, the actual variation 
  curve of this field moment must show significant deviations from the
  simple mathematical approximations used here. This in turn implies
  that the estimate of the rotation period that is derived by fitting
  such a simple mathematical curve to the observational data must be
  critically assessed for fine adjustment and determination of its
  uncertainty.

The time elapsed between \citeauthor{1958ApJS....3..141B}'s
\citeyearpar{1958ApJS....3..141B} first determination of the mean
longitudinal magnetic field of HD~50169 and our most recent
spectropolarimetric observation of the star is 23,500\,d, or $\sim$2.2
rotation periods. This provides a very sensitive approach to refine
the determination of the period and to estimate its uncertainty. By
plotting a phase diagram of the $\Hz$ measurements for a series of
tentative values of the period around the one suggested by the
periodogram, one can visually identify the period value that minimises
the phase shifts between field determinations from different rotation
cycles, and constrain the range around that value for which those
phase shifts remain reasonably small. More specifically, in the
present case, the ascending branch of the $\Hz$ variation curve proves
particularly well suited to the application of this method: the
measurements from \citet[][-- red crosses]{1958ApJS....3..141B} and our
$\Hz$ determinations from 6-m telescope observations (salmon-coloured filled
squares) should remain aligned along the same line, without
significant systematic phase shift of one set with respect to the
other. Critically, the accuracy of the period determination
  carried out in this manner only depends on the reproducibility of
  the variation curve from one cycle to the next, not on its exact
  shape.

Admittedly, $\Hz$ determinations obtained with different
instruments and by application of different data analysis methods are
known to show frequently systematic differences, whose existence
between the two sets of data of interest cannot be definitely ruled
out. However, previous studies indicate that, in general, for Ap stars
with fairly sharp spectral lines, there are at most minor
systematic differences between the longitudinal field values
determined, through application of the metal line spectropolarimetric
technique,    by Babcock (from observations obtained with the Mount 
Wilson 100-inch and Palomar 200-inch coud\'e spectrographs), by Mathys
and collaborators (from ESO CASPEC spectra), and by  Romanyuk and
collaborators (with the Main Stellar Spectrograph of the 6-m telescope
BTA of the Special Astrophysical Observatory). Assuming that the $\Hz$
data sets that we combine here are indeed mutually consistent
within their formal precision, we
determine the following best value of the rotation period of HD~50169:
\begin{equation}
  \Prot=(10,600\pm300)\,\mathrm{d}.
  \label{eq:Prot}
  \end{equation}

\begin{figure}
\resizebox{\hsize}{!}{\includegraphics[angle=0]{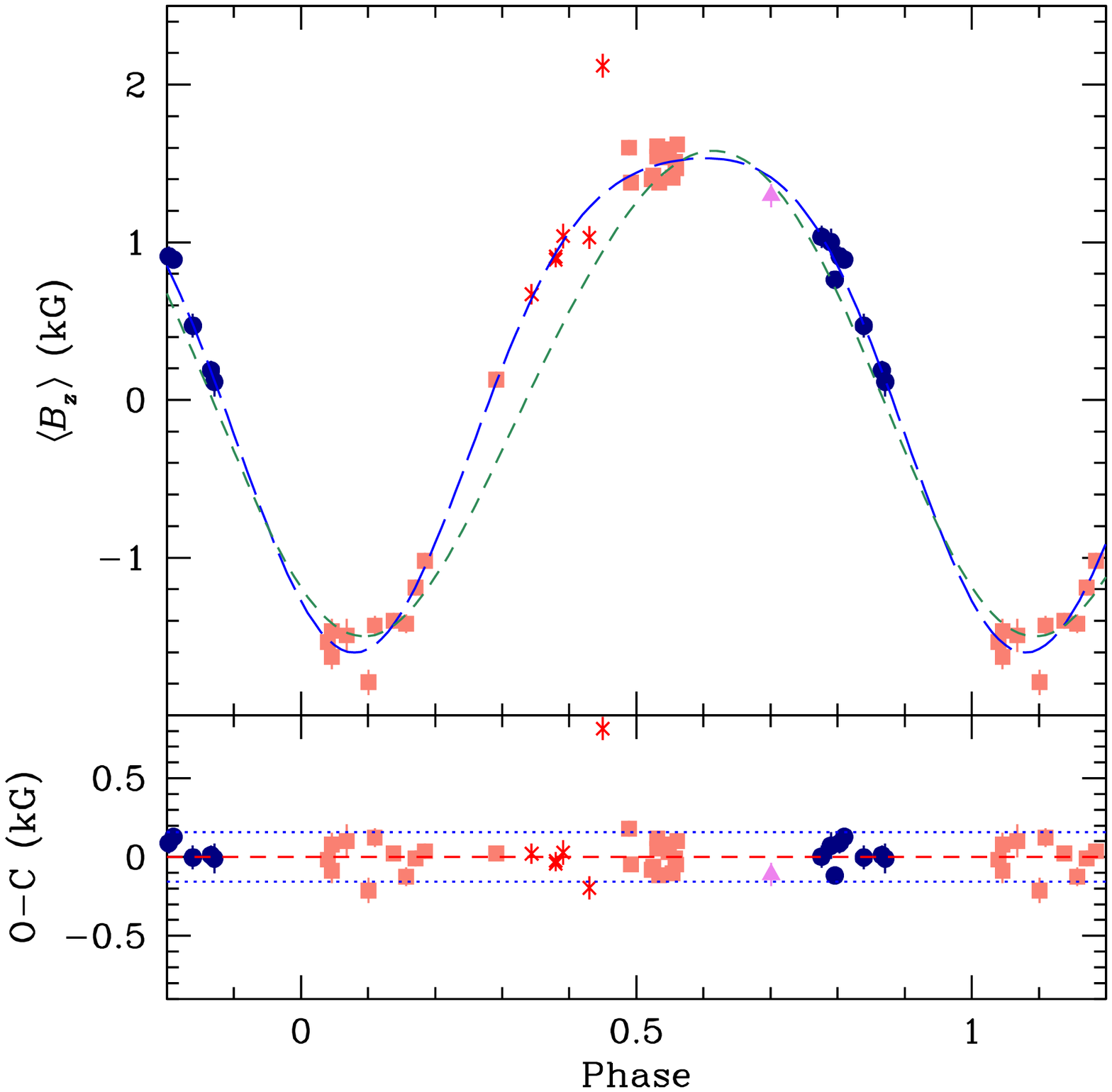}}
\caption{\textit{Upper panel:\/} Mean longitudinal magnetic field of
  HD~50169 against rotation 
  phase. The different symbols identify the source and/or instrumental
  configuration from which the $\Hz$ value was obtained, as follows:
  crosses (red): \citet{1958ApJS....3..141B}; filled triangle
  (violet): \citet{1997A&AS..124..475M}; filled circles (dark blue): 
  \citet{2017A&A...601A..14M}; filled squares (salmon):
  \citet{2014AstBu..69..427R}, and new measurements of spectra
  obtained with the Main Stellar Spectrograph of the 6-m telescope BTA
  of the Special Astrophysical Observatory. The
long-dashed line (blue) is the best fit of the observations by a cosine wave
and its first harmonic -- see Eq.~(\ref{eq:bzfit}). The short-dashed
line (green) corresponds to the superposition of low-order multipoles
discussed in Sect.~\ref{sec:magfield}. \textit{Lower panel:\/}
Differences $\mathrm{O}-\mathrm{C}$ between the individual $\Hz$
measurements and the best fit curve, against rotation phase. The
dotted lines (blue) correspond to $\pm1$~rms deviation of the
observational data about the fit (red dashed line). The symbols are
the same as in the upper panel.}
\label{fig:bzcurve}
\end{figure}

The phase variation curve of $\Hz$ for this value of the period is
shown in Fig.~\ref{fig:bzcurve}. Note the consistent behaviour of the
measurements of Babcock, and of Romanyuk's group, on the ascending
branch between phases 0.1 and 0.5. The estimated uncertainty
  on the period factors in possible systematic shifts between the
  measurement sets from the different groups that are at most of the
  same order of magnitude as the formal errors of these
  measurements. That is, we regard as unacceptably large, or
    unacceptably small, period values for which, in the phase
    diagram, within a given phase interval, a set of $\Hz$ data
    obtained by one of the groups is systematically offset with
    respect to the measurements by another group, performed in another
    rotation 
    cycle, by an amount that significantly exceeds the uncertainties
    of the individual $\Hz$ determinations. Admittedly, there is a
    certain degree of arbitrariness involved in this procedure, not
    only because it is based on a visual inspection of the variation
    curves, but also, more critically, because it relies on the
    assumption that the systematic differences between measurements
    obtained by different groups with different instruments do not
    exceed the random errors affecting these measurements. This
    assumption is borne out by our experience of the consistency of 
    the $\Hz$ measurements by the three groups in question in other
    stars. But that we cannot know for sure whether, for HD~50169,  there are
    sysematic differences between the measurements of the various
    groups and how large these differences are, implies that we cannot
    either determine in a fully objective and unambiguous way the
    uncertainty affecting the derived value of the period. The 300\,d
    value that we adopt for this uncertainty is our best, educated
    estimate, significantly higher than the uncertainty
      derived from a formal fit, 88\,d, which we regard as too low
      according to the arguments given above. Only the acquisition of
    additional measurements over a 
    large fraction of the next rotation cycle -- more than a decade --
    will allow the determination of the period and of its uncertainty
    to be further improved.

\begin{figure}
\resizebox{\hsize}{!}{\includegraphics[angle=0]{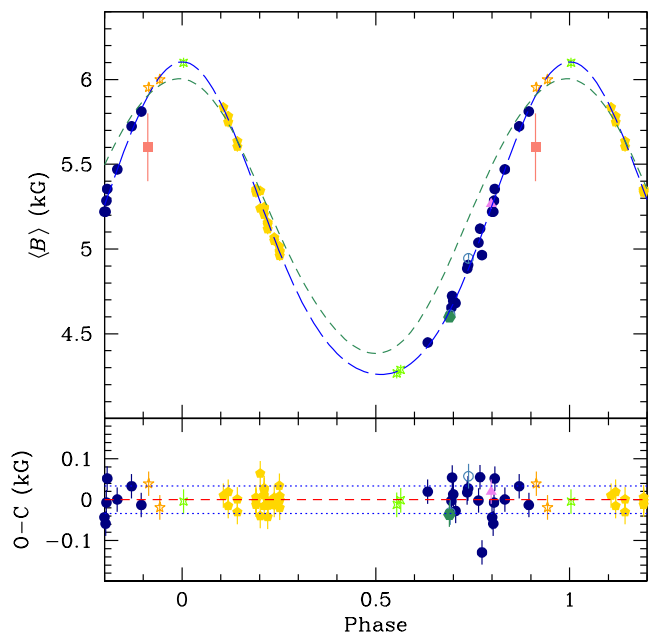}}
\caption{\textit{Upper panel:\/} Mean
  magnetic field modulus of HD~50169 against rotation
  phase. The different symbols identify the source and/or instrumental
  configuration from which the $\Hm$ value was obtained, as follows:
  filled circles (dark blue): CAT + CES LC; open circles (steel blue):
  CAT + CES SC; filled hexagon (sea green): 3.6\,m + CES LC; filled
  triangle (violet): CFHT + Gecko \citep[all previous 
  from][]{1997A&AS..123..353M,2017A&A...601A..14M}; five-pointed open stars
  (orange): 3.6\,m + CES VLC; four-pointed open stars (light green):
  UT2 + UVES; filled pentagons (yellow): 3.6\,m + HARPS; filled square
  (salmon): \citet{1971ApJ...164..309P}. The
long-dashed line (blue) is the best fit of the observations by a cosine wave
and its first harmonic -- see Eq.~(\ref{eq:bmfit}). The short-dashed
line (green) corresponds to the superposition of low-order multipoles
discussed in Sect.~\ref{sec:magfield}.  \textit{Lower panel:\/}
Differences $\mathrm{O}-\mathrm{C}$ between the individual $\Hm$
measurements and the best fit curve, against rotation phase. The
dotted lines (blue) correspond to $\pm1$~rms deviation of the
observational data about the fit (red dashed line). The symbols are
the same as in the upper panel.}
\label{fig:bmcurve}
\end{figure}

The 10,600\,d period value is also consistent with the observed
behaviour of the mean magnetic field modulus. Our measurements of this
field moment cover a shorter time span, 9856\,d. But they definitely
rule out a period shorter than that time span. The phase diagram of
the $\Hm$ variation for a period of 10,600\,d is shown in
Fig.~\ref{fig:bmcurve}. The way in which the measurement points line
up along
a smooth curve, with almost no scatter at all, is remarkable. It is
fully consistent with the estimated error of the $\Hm$
values. The error bars are actually plotted in the figure, but
as their size does not significantly exceed that of the symbols, they
can hardly be 
distinguished.

Moreover, agreement between the field values obtained
with the different instrumental configurations is excellent. We do not
find any systematic instrumental effects such as those that have
occasionally affected previous measurements \citep[e.g., Section
6 of][]{1997A&AS..123..353M}. In particular, as we used HARPS and UVES
spectra for the first time for $\Hm$ determinations, the good
agreement of the resulting values with those obtained with instruments
that we used in the past, such as the CES, is noteworthy. It 
strengthens our confidence that the mean magnetic field modulus
measurements that we obtain using a large number of different
instruments are free from any significant systematic errors of
instrumental origin.

As our $\Hm$ measurements do not cover yet a full rotation cycle, we
are not able to use the phase overlap between consecutive periods to
constrain the value of the period, contrary to what we did for the
mean longitudinal field. However, these measurements provide the
following useful indication. For values of the rotation period
comprised between 10,300 and 10,900\,d (that is, the uncertainty range
set from the longitudinal field analysis), the $\Hm$ variations can be
fitted well by a cosine curve and its first harmonic. This is quite
similar to the behaviour observed in the vast majority of the Ap stars
with resolved magnetically split lines. For instance, none of the
stars analysed by \citet{2017A&A...601A..14M} for which sufficient
phase coverage of the $\Hm$ measurements was achieved shows
significant deviations from the superposition of a cosine wave and its
first harmonic in the variations of this field moment. By contrast, if
we attempt to fit the variations of the mean magnetic field modulus of
HD~50169 with a period significantly outside the range
10,300--10,900\,d, more 
harmonics become required in order for the fitted curve to match the
observations.

Admittedly, we cannot definitely rule out the possibility of an
unusual structure of the magnetic field of HD~50169. But this
improbable coincidence appears all the less plausible since the
resolved magnetically split components of the \Feline\ line in this
star are particularly sharp and clean, while increased complexity in
the geometrical structure of the magnetic field tends to manifest
itself by distortions of the split line components, as was found
e.g. in a recent study of HD~18078 \citep{2016A&A...586A..85M}.

As already noted in the introduction, \citet{1971ApJ...164..309P} was
the first to obtain an estimate of the mean magnetic field modulus of
HD~50169: 5.6\,kG, derived from consideration of the differential
broadening of spectral lines having Zeeman patterns of different
widths. We retrieved the Julian Date of this observation from
\citet{1973ApJS...26....1A}, as the same spectrum was used for 
detailed abundance analysis. The representative point of
\citeauthor{1971ApJ...164..309P}'s field determination appears as a
salmon-coloured filled square in Fig.~\ref{fig:bmcurve}.

The uncertainty of this determination is difficult to assess. The most
meaningful comparison between the field values derived by
\citet{1971ApJ...164..309P} through application of his Equation~(3)
and the $\Hm$ determinations of \citet{2017A&A...601A..14M}, based on
the splitting of the \Feline\ doublet, is for those stars for which
the mean field modulus only shows small, almost insignificant
variations, such as HD~2453, HD~137949, and HD~192678. For these
stars, the estimates of \citeauthor{1971ApJ...164..309P} and the
measurements of  \citeauthor{2017A&A...601A..14M} are in
excellent agreement, within $\sim$100\,G of each other. The comparison 
is less straightforward when the field modulus shows definite
variability with phase, but at least for a couple of stars (e.g.,
HD~12288 or HD~188041), the agreement between the values that were
derived from line broadening by \citeauthor{1971ApJ...164..309P} and
those that  \citeauthor{2017A&A...601A..14M} obtained from
consideration of the wavelength separation of split line components,
seems somewhat poorer. The error bar on $\Hm$ that we adopted for the
data point from \citet{1971ApJ...164..309P} in Fig.~\ref{fig:bmcurve}, 
$\pm200$\,G, is only tentative, and it may well underestimate the
actual uncertainty of the field values reported in this
reference. But even with this ``optimistic'' accuracy, the magnetic
field strength estimated nearly 50 years ago by
\citeauthor{1971ApJ...164..309P} is consistent with the value of the
rotation period that we derived above.

With $\Prot=10,600$\,d, HD~50169 becomes the longest-period Ap star
for which magnetic field measurements have been obtained over more
than a full rotation cycle. At 29.04\,y, its period is considerably
longer than the next longest one, 21.75\,y for HD~9996. Furthermore,
with respect to the latter, HD~50169 presents the advantage of
showing resolved magnetically split lines throughout its whole rotation
cycle, so that its mean magnetic field modulus can be exactly
determined even around the phase of its minimum and, as a consequence,
the structure of the field can be better constrained. 

\section{Magnetic field characterisation}
\label{sec:magfield}
The observed variations of both the mean longitudinal magnetic field
and the mean magnetic field modulus of HD~50169 can be well
represented by the superposition of a cosine wave and of its first
harmonic. The best least-squares fit solutions for $\Prot=10,600$\,d
are:
\begin{eqnarray}
\Hz(\phi)&=&(253\pm30)\nonumber\\
&+&(1566\pm50)\,\cos\{2\pi\,[\phi-(0.583\pm0.005)]\}\nonumber\\
         &+&(290\pm44)\,\cos\{2\pi\,[2\phi-(0.651\pm0.027)]\}\nonumber\\
         &&\hspace{-3em}(\nu=34,\ \chi^2/\nu=9.8),\label{eq:bzfit}\\
  \nonumber\\
\Hm(\phi)&=&(5076\pm10)\nonumber\\
&+&(922\pm15)\,\cos\{2\pi\,[\phi-(0.004\pm0.001)]\}\nonumber\\
         &+&(107\pm10)\,\cos\{2\pi\,[2\phi-(0.988\pm0.017)]\}\nonumber\\
  &&\hspace{-3em}(\nu=44,\ \chi^2/\nu=1.4),\label{eq:bmfit}
\end{eqnarray}
where the field strengths are expressed in Gauss, $\phi=({\rm
  HJD}-{\rm HJD}_0)/\Prot$ (mod 1) and the adopted 
value of ${\rm HJD}_0=2441600.0$ corresponds to a maximum of the mean
magnetic field modulus; $\nu$ is the number of degrees of
  freedom, and $\chi^2/\nu$, the reduced $\chi^2$ of the fit. The
fitted curves 
are shown in Figs.~\ref{fig:bzcurve} and \ref{fig:bmcurve};
the $\mathrm{O}-\mathrm{C}$ differences between the individual
  measurements and these curves are also illustrated. The $\Hm$
estimate of \cite{1971ApJ...164..309P} was not taken into account for
computation of the best fit. Its inclusion would only have affected
the derived fit parameters insignificantly. The quality of the
 mean longitudinal magnetic field  measurement on JD\,2435765 is also
  debatable, as it appears to be a definite outlier. In the absence of
  any comment from \citet{1958ApJS....3..141B} about it, we included
  it in the $\Hz$ fit. Omitting it would not have yielded
  significantly different fit parameters, but the reduced $\chi^2$ would
  have been decreased from 9.8 to 6.3 (for 33 degrees of freedom). That this
  value is still rather high may indicate that the uncertainty of
  some of the measurements is slightly underestimated, that there are
  some (small) systematic differences between the measurements of the
  different groups, or that the actual shape of the $\Hz$ variation
  curve departs somewhat from the superposition of a cosine wave and
  its first harmonic -- or some combination of these effects. This
  further strengthens our conviction that our conservative approach for
  the determination of the rotation period and of its uncertainty,
  based on the visual inspection of the consistency of measurements
  obtained at similar phases in different cycles and making allowance
  for unidentified (small) systematic errors, is warranted.

Within the estimated $\pm300$\,d uncertainty in the derived value of
the rotation period, the $\Hm$ variation curve  can be more or less
stretched, but its general shape remains unchanged, mostly symmetrical
about the phases of both its maximum ($\phi_{\rm M,max}=0.0$) and its
minimum ($\phi_{\rm M,min}=0.5$). The magnetic minimum is somewhat
broader and shallower than the maximum, as is frequently the case
\citep{2017A&A...601A..14M}. 

Because the mean longitudinal field measurements span more than two
rotation cycles, the $\Hz$ variation curve is more sensitive to the
length of the rotation period. In particular, depending on the exact
value of this period, the phases of the extrema of $\Hz$ may be more
or less shifted with respect to those of $\Hm$, and the $\Hz$
variation curve may show more or less pronounced -- although never
large -- departures from symmetry about those phases. Regardless, the
main features of the variations of the longitudinal field are quite
definite: the phases of the extrema of $\Hz$ lag significantly behind
those of $\Hm$, by a (period-dependent) amount close to 0.1 rotation
cycle, and the $\Hm$ maximum occurs close to the negative extremum of
$\Hz$, indicating that the magnetic field is stronger in the vicinity
of the negative pole of the star than around the positive pole. The positive
extremum of the longitudinal field is broader and shallower than the
negative one, mirroring the above-mentioned difference between the
minimum and the maximum of the $\Hm$ curve.

Simple parametrised magnetic field models of the distribution of
magnetic flux over the stellar surface are frequently fitted to
magnetic moment curves such as those of Figs.~\ref{fig:bzcurve} and
\ref{fig:bmcurve} in order to 
obtain a crude idea of the large-scale magnetic field structure of a
star. These models, typically requiring only four or five parameters,
certainly do not provide realistic details of the distribution of
magnetic flux over the surface of the star, but they do allow us to
estimate such interesting geometric parameters as the inclination $i$
of the rotation axis to the line of sight, and the obliquity $\beta$
of the magnetic axis (supposing that the field is nearly enough
axisymmetric to have such an axis) to the rotation axis.

Amongst the simplest of such models is an expansion of the surface
field in low order axisymmetric multipoles \citep{1950MNRAS.110..395S},
using 
only the first two or three terms, for example using collinear dipole,
quadrupole, and octupole \citep{1988ApJ...326..967L}. Such models can
usually 
be found which fit the available magnetic moment measurements
reasonably well \citep[e.g.][]{2000A&A...359..213L}. When such models
are fit to a sample of magnetic Ap stars, a distinctive feature
generally found is that the dipole component is almost always a major
contributor to the expansion, while the importance of the higher terms
is quite variable. A basic symptom of an important dipole component is
that the largest ratio of $|\Hz/\Hm|$
is of the order of 0.3. This is the case for HD\,50169. 

However, for this star it is obvious that the field distribution is
{\em not} axisymmetric. An axisymmetric field distribution has the
property that there are two phases, separated by 0.5 phase, through
which all the field moment curves are reflectionally symmetric. The
phase shift between the extrema of the $\Hz$ and
$\Hm$ curves guarantees that there are no such phase
pairs. Hence we must look for a simple field geometry that lacks
axisymmetry.

\begin{figure}
\resizebox{\hsize}{!}{\includegraphics{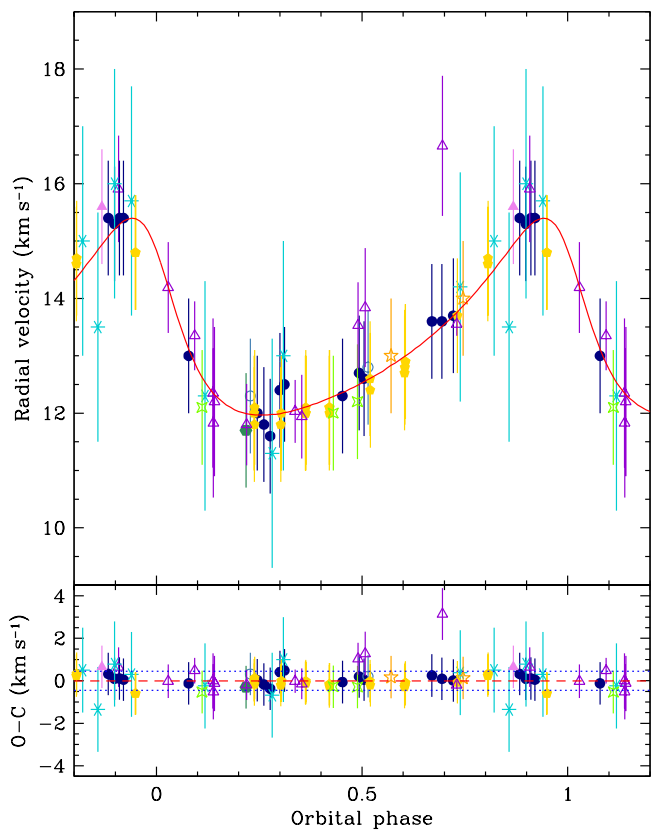}}
\caption{{\it Upper panel\/}: Radial velocity measurements for
  HD~50169, plotted against orbital phase. The solid curve 
  corresponds to the orbital solution given in
  Sect.~\ref{sec:bin}. The time $T_0$ of periastron passage is
  adopted as phase origin. {\it Bottom panel:\/} Plot of the
  differences ${\rm O}-{\rm C}$ between the observed values of the
  radial velocity and the predicted values computed from the orbital
  solution. The  dotted lines correspond to $\pm1$ rms
  deviation of the observational data about the orbital solution 
    (dashed line). Open triangles (dark violet) represent data from
    \citet{2002A&A...394..151C} and  asterisks (turquoise) ESO CASPEC 
  observations from \citet{1997A&AS..124..475M} and
  \citet{2017A&A...601A..14M}; all other symbols have the same
  meaning as in Fig.~\ref{fig:bmcurve}.}  
\label{fig:rv}
\end{figure}

In principle, if we allow enough free parameters (for example, by
taking an expansion in dipole, quadrupole, and octupole, all
independently orientated with respect to the stellar rotation axis), we
could probably fit the details of the observed field moment curves
quite accurately, but also quite possibly non-uniquely. Instead, we
look for a simple single parameter that we could add to the usual
three coaxial multipole components, that reproduces approximately the
phase shift between the extrema, without trying to fit all
details of the (extremely well defined) field moment curves.

A simple alteration to the usual axisymmetric multipole model that
breaks the axisymmetry of the field model, and is easy to program, is
to introduce a gradient in field strength in the direction normal to
the plane containing the rotation and multipole axes. For example,
taking a Cartesian coordinate system with the plane of the rotation
and multipole axes as the $x-z$ plane, we can simply multiply all
field strengths at a given $y$--distance from the $x-z$ plane by a
factor $1 + Ay$, where $A < 1$ is a single new parameter. This choice
preserves $\mathrm{div}\,\vec{B} = 0$, while making the field somewhat
stronger 
in one magnetic hemisphere than in the other. This choice also is
found to lead to a shift in the relative extrema of one moment curve
relative to the other that is approximately proportional to $A$.

Using a simple {\sc fortran} program that computes the field moments
of a given field distribution, and searches for the multipole
parameter values that agree best with a set of samples of the moments
at various phases, we find that the best parameter values are $i =
40^\circ$, $\beta = 92^\circ$ (note that $i$ and $\beta$ may be
exchanged with no change in the predicted curves), $B_{\rm dipole} =
-6930$\,G, $B_{\rm quadrupole} = -2490$\,G, $B_{\rm octupole} =
+2880$\,G, and $A = 0.14$. It will be seen from Figs.~\ref{fig:bzcurve} and
\ref{fig:bmcurve} that the
model computed with these parameters fits the observed moments fairly
well, and in particular it does produce about the observed phase shift
between the extrema. As expected, the dipole is the dominating
component of the field structure, but an asymmetry such as that
introduced by $A \neq 0$ is essential for the phase shift.

\section{Binarity}
\label{sec:bin}
\citet{1997A&AS..123..353M} were the first to report the variability
of the radial velocity of HD~50169. They suggested that the star could
be a binary. This suspicion was supported by independent,
contemporaneous radial 
velocity measurements from \cite{2002A&A...394..151C}. Combining the
latter with an extensive set of data built up as a by-product of his
systematic study of the magnetic field of the star,
\cite{2017A&A...601A..14M} computed for the first time the orbital
elements of this binary. With the addition of the new radial velocity
measurements obtained as part of the present study, the time base is
increased by a factor of 3.6, albeit with a less dense sampling. This
enables us to refine considerably the definition of most orbital
elements, as follows:
\begin{itemize}
\item[]$P_{\rm orb}=(1757.6\pm17.3)$\,d
\item[]$T_0=$HJD\,$(2,448,532\pm39)$
\item[]$e=0.384\pm0.059$
\item[]$V_0=(13.24\pm0.07)$\,km\,s$^{-1}$
\item[]$\omega=47\fdg7\pm9\fdg2$
\item[]$K=(1.72\pm0.13)$\,km\,s$^{-1}$
\item[]$f(M)=(0.0007\pm0.0073)$\, $M_{\sun}$
\item[]$a\,\sin i=(38.3\pm3.1)$\, $10^6$\,km
\end{itemize}
This solution, which was computed using the Li\`ege Orbital Solution
Package\footnote{\texttt{http://www.stsci.edu/\textasciitilde{}hsana/losp.html}}
(LOSP), is illustrated in Fig.~\ref{fig:rv}.

The mass function is exceptionally small. For instance, among the Ap
binaries for which the orbital elements are listed by
\citet{2002A&A...394..151C} or by \citet{2017A&A...601A..14M}, only
HD~200405 has a lower mass function, $f(M)=0.00010$. However, this
most likely results from the low inclination of the orbital plane,
based on the (different) arguments given in both studies.

By contrast, if we assume that the axis of rotation of the Ap
component of HD~50169 is roughly perpendicular to the orbital plane,
the inclination of the latter must be rather large, since we found
in Sect.~\ref{sec:magfield} that $i$ is of the order of either
$40^\circ$ or $92^\circ$. Thus, the
low value of $f(M)$ must reflect the low mass of the
secondary.

Assuming that the mass of the Ap star is of the order of
2\,$M_\sun$, for $i=40^\circ$, we derive a $1\sigma$ upper limit of
$\sim$0.55\,$M_\sun$ for the mass of its companion. For $i=92^\circ$,
this upper limit is reduced to $\sim$0.35\,$M_\sun$. These upper
limits correspond, respectively, spectral types K7--M1 or M2--M3 (on the
main sequence). In either case, the secondary of HD~50169 would
certainly be one of the latest-type companions known of any Ap
star. It might possibly even be a late M or early L star, given the
uncertainty on the mass function. 

\section{Discussion}
\label{sec:concl}
Since the existence of Ap stars with unexpectedly long rotation
periods became apparent, and as more were progressively found,
some of them with ever longer periods, the question has repeatedly
arisen whether there were other differences distinguishing
them from their faster rotating counterparts. This question is
relevant not only to decide whether the short- and long-period Ap
stars constitute two different classes of objects, or if they belong
to a single, mostly uniform group. Answering it is also potentially
important to shed light onto the physical mechanisms that must be at
play to achieve extremely slow rotation, or more generally, the huge
differentiation of rotation rates among Ap stars.

In his extensive study of the magnetic properties of Ap stars with
resolved magnetically split lines, \citet{2017A&A...601A..14M} used a
number of numerical parameters to characterise the behaviour of the
sample stars and to investigate possible trends among them. HD~50169
is now the longest-period star for which these parameters are fully
determined. Their consideration provides the first opportunity to test
how different HD~50169 is from less slowly rotating Ap stars, or how
similar to them.

The average value of the mean magnetic field modulus of HD~50169 over
its rotation 
period is $\Hav=5181$\,G. This is fully consistent with the trend
illustrated in Figure~2 of \citet{2017A&A...601A..14M}, according to
which magnetic fields in excess of 7.5\,kG are found only in Ap stars
with rotation periods shorter than 150\,d. The ratio between the
extrema of the mean magnetic field modulus, $q=1.43$, unambiguously
puts the representative point of HD~50169 in the upper right quadrant
of Figure~4 
of \citet{2017A&A...601A..14M}. This strengthens the suspicion that
the relative amplitude of the $\Hm$ variations tends to be greater in
longer period stars. The degree of anharmonicity of the $\Hm$
variation curve in HD~50169 is well within the range observed in other
stars with periods of several years \citep[see Figure~5
of][]{2017A&A...601A..14M}. The ratio of the root-mean-square (rms)
longitudinal field \citep{1993A&A...269..355B}, $\Hzrms=1294$\,G, to
$\Hav$ is 0.25, also well within the typical range \citep[see Figure~8
of][]{2017A&A...601A..14M}. The combination of $q=1.43$ with a value
$r=-0.96$ of the ratio of the smaller (in absolute
value) to the larger (in absolute value) extremum of $\Hz$ is
consistent with the idea that large relative amplitudes of variation 
of the field modulus are observed in stars for which both poles come
alternatively into sight \citep[see Figure~9
of][]{2017A&A...601A..14M}. The phase difference (0.579) between the fundamentals
of the fits of the measurements of $\Hz$ and $\Hm$ by functions of the
forms given in Eqs.~(\ref{eq:bzfit}) and (\ref{eq:bmfit}) is well
within one of the ``normality bands'' of Figure~11 of
\citet{2017A&A...601A..14M}.

In summary, in all respects, HD~50169 is a ``well-behaved'' Ap star,
which does not exhibit any of the ``anomalies'' found in
several of the stars studied by \citet{2017A&A...601A..14M}. This is
consistent with the view that there is no fundamental difference
distinguishing the longest period Ap stars from the rest of the class,
or in other words, that the Ap stars constitute a single unified class
of objects spanning an exceptionally broad range of rotation rates, but
otherwise having similar physical properties. However, this does not
imply that there is no correlation between the rotation periods and
other properties of Ap stars. The apparent dichotomy in the
distribution of the magnetic field strengths between stars with
periods shorter and longer than 150 days reported by
\citet{2017A&A...601A..14M} is an example of such a correlation, and
this author discussed other possible connections between e.g. the
geometric structure of the magnetic field and the rotation period. But
such correlations can only be evidenced and confirmed through studies
of statistically significant samples of stars.

At present, this statistical significance is at least partially
achieved over three orders of magnitude, from $10^0$ to a few
$10^2$\,d. Extending it to higher orders of magnitude represents a
daunting challenge. The first stellar magnetic field was discovered
$\sim$2.5 rotation periods of HD~50169 ago; $\sim$1.7 such period
elapsed 
since the interest of the long-period Ap stars was recognised; and
less than one period of this star has been completed since systematic
searches for 
Ap stars with resolved magnetically split lines have started. On these
timescales, the difference ($\sim$7\,y) between its period and that of
HD~9996, which was previously the longest-period Ap star for which
magnetic measurements had been obtained over a full rotation cycle,
represents a considerable step. But in the broader context of
the five orders of magnitude (or more) spanned by the periods of the
Ap stars, it is hardly meaningful.

The progress that can be achieved
towards improved knowledge of the most slowly rotating Ap stars is by
nature incremental, since the only way to increase the time base over
which relevant observations are available is to await the passage of
time. Most likely, the star that will someday overturn HD~50169 as the
longest-period Ap star with full coverage of the magnetic variation
curve will only have a rotation period a few years, or possibly a
couple of decades, longer. This does not detract from the scientific
value of 
monitoring the magnetic fields of the extremely slowly rotating Ap
stars. On the contrary, we owe it to the next generations to make sure
that we continue to collect the relevant data on a sufficiently
regular basis, avoiding gaps in the time series that may severely
jeopardise future analyses, so that eventually our successors can
successfully achieve complete knowledge of those fascinating stars
that hardly rotate at all. The legacy of previous generations opens
the realistic prospect of achieving in our lifetime full knowledge of
a star such as $\gamma$~Equ \citep[see][and references
therein]{2016MNRAS.455.2567B}. That this star is no longer unique, and
that there must exist instead a significant population of Ap stars
with periods of the order of centuries, does not exonerate us from the
responsibility of documenting the behaviour of these stars. If we
failed to fulfil this responsibility for continuity, our carelessness
would deserve to be condemned by our descendants.

\begin{acknowledgements}
We are grateful to Drs. G. W. Preston and S. J. Adelman for their help
in retrieving the date and time at which the spectrum from which
\cite{1971ApJ...164..309P} estimated the mean magnetic field modulus
of HD~50169 was acquired. Parts of this study were carried out during
a stay of GM in the Department of Physics \& Astronomy of the
University of Western Ontario (London, Ontario, Canada) funded by the
ESO Science Support Discretionary Fund (SSDF), and a stay of SH at the
ESO office in Santiago within the framework of the ESO Santiago
visitors programme. Thanks are due to ESO for its financial support of
these science stays, and to the respective host institutions for their
welcome. IIR (RSF grant No. 18-12-00423), and DOK, EAS and IAY (RSF
grant No. 14-50-00043) gratefully acknowledge
the Russian Science Foundation for partial financial support.
This research has made use of the SIMBAD
database, operated at the CDS, Strasbourg, France.
\end{acknowledgements}

\bibliographystyle{aa}
\bibliography{hd50169}
\end{document}